\documentclass[prl,twocolumn]{revtex4}

\usepackage{graphicx}
\usepackage{dcolumn}
\usepackage{bm}

\newcommand{\Li}{\mathrm{Li}}
\newcommand{\mean}[1]{\langle #1 \rangle}

\begin{document}

\preprint{Preprint}
\bibliographystyle{prsty}

\title{Information Flow in Social Groups}
\author{Fang Wu, Bernardo A. Huberman, Lada A. Adamic and Joshua R. Tyler}
\affiliation{Information Dynamics Lab, HP Laboratories, 1501 Page
Mill Road, CA 94304-1126}

\begin{abstract}
We present a study of information flow that takes into account the
observation that an item relevant to one person is more likely to
be of interest to individuals in the same social circle than those
outside of it. This is due to the fact that the similarity of node
attributes in social networks decreases as a function of the graph
distance. An epidemic model on a scale-free network with this
property has a finite threshold, implying that the spread of
information is limited. We tested our predictions by measuring the
spread of messages in an organization and also by numerical
experiments that take into consideration the organizational
distance among individuals.
\end{abstract}

\pacs{epidemics, scale-free networks, information flow}
\maketitle

The problem of information flows in social organizations is
relevant to issues of productivity, innovation and the sorting out
of useful ideas out of the general chatter of a community. How
information spreads determines the speed with which individuals
can act and plan their future activities. In particular, email has
become the predominant means of communication in the information
society. It pervades business, social and scientific exchanges and
as such it is a highly relevant area for research on communities
and social networks. Not surprisingly, email has been established
as an indicator of collaboration and knowledge exchange
\cite{wellman02computersocial, whittaker96email, guimera02email,
tyler03email,eckmann03email}. Email is also a good medium for
research because it provides plentiful data on personal
communication in an electronic form.

Since individuals tend to organize both formally and informally
into groups based on their common activities and interests, the
way information spreads is affected by the topology of the
interaction network, not unlike the spread of a disease among
individuals. Thus one would expect that epidemic models on graphs
are relevant to the study of information flow in organizations. In
particular, recent work on epidemic propagation on scale free
networks found that the threshold for an epidemic is zero,
implying that a finite fraction of the graph becomes infected for
arbitrarily low transmission probabilities
\cite{zoltan02halting,pastor-satorras01epidemic,newman02emailnetworks}.
The presence of additional network structure was found to further
influence the spread of disease on scale-free graphs
\cite{eguiluz02epidemicclust,vazquezPRE2003,newman02assortative}.

There are, however, differences between information flows and the
spread of viruses. While viruses tend to be indiscriminate,
infecting any susceptible individual, information is selective and
passed by its host only to individuals the host thinks would be
interested in it.

\begin{figure}[tbp]
\begin{center}
\includegraphics[scale=0.4]{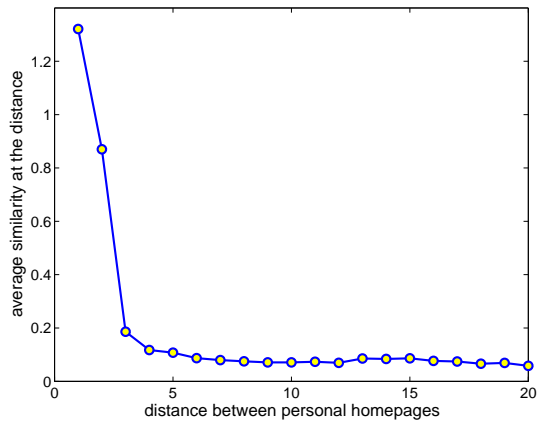}
\end{center}
\caption[Node similarity as a function of distance in the graph
]{Average similarity of Stanford student homepages as a function
of the number of hyperlinks separating them.} \label{disttolikeav}
\end{figure}

The information any individual is interested in depends strongly
on their characteristics. Furthermore, individuals with similar
characteristics tend to associate with one another, a phenomenon
known as homophily \cite{lazarsfeld54friendship,
touhey74similarity,feld81social}. Conversely, individuals many
steps removed in a social network on average tend not to have as
much in common, as shown in a study \cite{adamic03friends} of a
network of Stanford student homepages and illustrated in Figure
\ref{disttolikeav}.

We therefore introduce an epidemic model with decay in the
transmission probability of a particular piece of information as a
function of the distance between the originating source and the
current potential target. In the following analysis, we show that
this epidemic model on a scale-free network has a finite
threshold, implying that the spread of information is limited. We
further tested our predictions by observing the prevalence of
messages in an organization and also by numerical experiments that
take into consideration the organizational distance among
individuals.

Consider the problem of information transmission in a power-law
network whose degree distribution is given by
\cite{newman01graphs} \begin{equation} p_k =
Ck^{-\alpha},\end{equation} where $\alpha>1$, and $C$ is
determined by the normalization condition. The generating function
of the distribution is \begin{equation} G_0(x) = \sum_{k=0}^\infty
p_k x^k = \frac{\Li_\alpha(x)}{\Li_\alpha(1)}.
\end{equation}

Following the analysis in \cite{newman02epidemic} for the SIR
(susceptible, infected, removed) model, we now estimate the
probability $p_m^{(1)}$ that the first person in the community who
has received a piece of information will transmit it to $m$ of
their neighbors. Using the binomial distribution, we find
\begin{equation} p_m^{(1)} = \sum_{k=m}^\infty p_k {k \choose m}
T^m (1-T)^{k-m},
\end{equation} where the \emph{transmissiblity} $T$ is the probability that a person
will transmit an item to a neighbor and the superscript ``$(1)$''
refers to first neighbors, those who received the information
directly from the initial source. The generating function for
$p_m^{(1)}$ is given by
\begin{eqnarray}
G^{(1)} (x) &=& \sum_{m=0}^{\infty} \sum_{k=m}^\infty p_k {k
\choose m} T^m (1-T)^{k-m} x^m\\
&=& G_0(1+(x-1)T) = G_0(x;T).
\end{eqnarray}

Suppose the transmissibility decays as a power of the distance
from the initial source. We choose this weakest form of decay as
the results that are obtained from it will also be valid for
stronger functional forms. Then the probability that an $m$th
neighbor will transmit the information to a person with whom he
has contact is given by \begin{equation} T^{(m)} =
(m+1)^{-\beta}T,\end{equation}
 where $\beta>0$ is the decay constant. $T^{(m)}=T$ at the originating node ($m=0$)
 and decays to zero as $m\rightarrow\infty$.

The distribution of the number of 2nd neighbors can be written as
\begin{equation} G^{(2)}(x) = \sum_k p_k^{(1)} [G_1^{(1)}(x)]^k = G^{(1)}
(G_1^{(1)}(x)), \end{equation} where \begin{equation} G_1^{(1)}(x)
= G_1(x; 2^{-\beta}T) = G_1(1+ (x-1) 2^{-\beta}T). \end{equation}
Similarly, if we define $G^{(m)}(x)$ to be the the generating
function for the number of $m$th neighbors affected, then we have
\begin{equation} G^{(m+1)}(x) = G^{(m)} (G_1^{(m)}(x)) \quad
\mbox{for }m\ge 1, \end{equation} where
\begin{equation} G_1^{(m)}(x) = G_1(x; (m+1)^{-\beta}T) = G_1(1+
(x-1)(m+1)^{-\beta}T), \end{equation} and
\begin{equation}G_1(x) = \frac{G_0'(x)}{G_0'(1)}=\frac 1z G_0'(x).\end{equation} Or, more explicitly, \begin{equation} G^{(m+1)}(x) = G^{(1)} ( G_1^{(1)}
( G_1^{(2)} ( \cdots G_1^{(m)} (x)))). \end{equation}

The average number $z_{m+1}$ of $(m+1)$th neighbors is
\begin{equation} z_{m+1} = {G^{(m+1)}}'(1) = {G_1^{(m)}}'(1)
{G^{(m)}}'(1) = {G_1^{(m)}}'(1) z_m. \end{equation} So the
condition that the size of the outbreak remains finite is given by
\begin{equation} \frac{z_{m+1}} {z_m} = {G_1^{(m)}}'(1) < 1, \end{equation} or
\begin{equation} (m+1)^{-\beta}T G_1'(1) < 1.
\end{equation} For any given $T$, the left hand side of the inequality above
goes to zero when $m\to \infty$, so the condition is eventually
satisfied for large $m$. Therefore the average total size
\begin{equation} \mean s = \sum_{m=1}^\infty z_m \end{equation} is always
finite if the transmissibility decays with distance. Note that if
$T$ is constant the average total size is infinite for values of
$\alpha < 3$ as shown previously.

\begin{figure}
  \centering\includegraphics[scale=0.45]{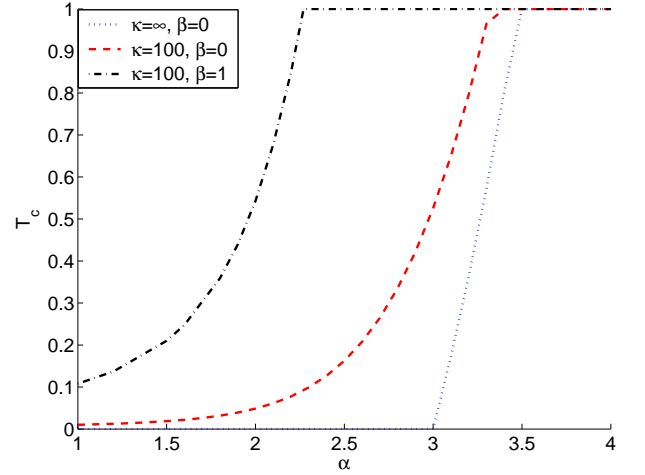}
  \caption{\label{threshold} $T_c$ as a function of $\alpha$. The three different curves,
  from bottom to top are: 1) no decay in transmission probability, no exponential cutoff in the
  degree distribution ($\kappa=\infty, \beta=0$). 2) $\kappa=100,
  \beta=0$, 3) $\kappa=100, \beta=1$.}
\end{figure}

In the real world however, the size of a network is always finite,
and in order to define a transmissibility threshold one needs an
outbreak size that is compatible with the size of the whole
network. Furthermore, many real world networks have a cutoff
$\kappa$ far below their size. Thus we can write for the link
distribution $ p_k = Ck^{-\alpha}\exp(-k/\kappa)$.

As an example, consider a network made up of $10^6$ vertices. We
define an epidemic to be an outbreak affecting more than 1\% or
$10^4$ vertices. Thus for fixed $\alpha, \kappa$ and $\beta$, we
can define $T_c$ as the transmissibility above which $\mean s$
would be made to exceed $10^4$.

The numerical result of $T_c$ as a function of $\alpha$ is shown
in Fig.~$\ref{threshold}$, where we choose $\kappa=100$ and
$\beta=1$. It is seen that when there is no decay, $T_c$ is very
near zero for $\alpha$ close to 2, which means that for most
values of $T$ epidemics occur. However, when the transmissibility
decays, $T_c$ rises substantially. For example, $T_c$ jumps to
0.54 at $\alpha=2$, implying that the information may not spread
over the network.

\begin{figure}[tbp]
\begin{center}
\includegraphics[scale=0.45]{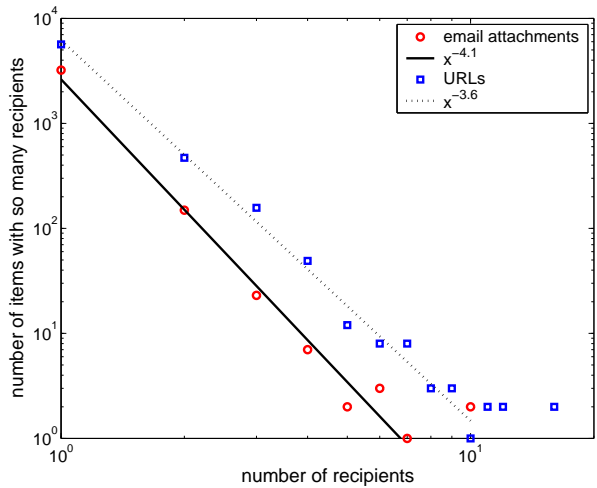}
\end{center}
\caption[Number of people receiving URLs and attachments]{Number
of people receiving URLs and attachments } \label{urlsattachments}
\end{figure}

In order to validate empirically that the spread of information
within a network of people is limited, and hence distinct from the
spread of a virus, we gathered a sample from the mail clients of
40 individuals (30 within HP Labs, and 10 from other areas of HP,
other research labs, and universities).  Each volunteer executed a
program that identified URLs and attachments in the messages in
their mailboxes, as well as they time the messages were received.
This data was cryptographically hashed to protect the privacy of
the users.  By analyzing the message content and headers, we
restricted our data to include only messages which had been
forwarded at least one time, thereby eliminating most postings to
mailing lists and more closely approximating true inter-personal
information spreading behavior. The median number of messages in a
mailbox in our sample is 2200, indicating that many users keep a
substantial portion of their email correspondence. Although some
messages may have been lost when users deleted them, we assume
that a majority of messages containing useful information had been
retained.

Figure \ref{urlsattachments} shows a histogram of how many users
had received each of the 3401 attachments and 6370 URLs. The
distribution shows that only a small fraction (5\% of attachments
and 10\% of URLs) reach more than 1 recipient. Very few (41 URLs
and 6 attachments) reached more than 5 individuals, a number
which, in a sample of 40, starts to resemble an outbreak. In
follow-up discussions with our study subjects, we were able to
identify the content and significance of most of these messages.
14 of the URLs were advertisements attached to the bottom of an
email by free email services such as Yahoo and MSN. These are in a
sense viral, because the sender is sending them involuntarily. It
is this viral strategy that was responsible for the rapid buildup
of the Hotmail free email service user base.  10 URLs pointed to
internal HP project or personal pages, 3 URLs were for external
commercial or personal sites, and the remaining 14 could not be
identified.

In our sample, one group is overrepresented, allowing us to
observe both the spread of information within a close group, and
the lack of information spread across groups. A number of
attachments reaching four or more people were resumes circulated
within one group. A few attachments were announcements passed down
by higher level management. This kind of top down transmission
within an organization is another path through which information
can be efficiently disseminated.

\begin{figure}[tbp]
\begin{center}
\includegraphics[scale=0.45]{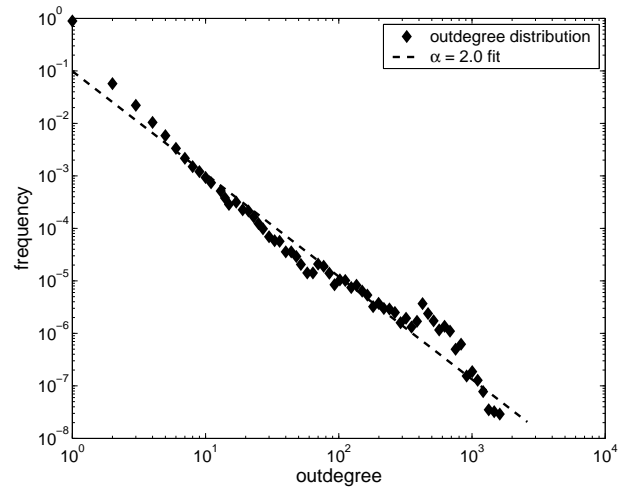}
\end{center}
\caption[Outdegree distribution for HP Labs email graph]{Outdegree
distribution for all senders (224,514 in total) sending email to
or from the HP Labs email server over the course of 3 months. The
outdegree of a node is the number of correspondents the node sent
email to.} \label{outdegdist}
\end{figure}

Next we simulated the effect of decay in the transmission
probability on the email graph at HP Labs in Palo Alto, CA. The
graph was constructed from recorded logs of all incoming and
outgoing messages over a period of 3 months.  The graph has a
nearly power-law out degree distribution, shown in Figure
\ref{outdegdist}, including both internal and external nodes.
Because all of the outgoing and incoming contacts were recorded
for internal nodes, their in and out degrees were higher than for
the external nodes for which we could only record the email they
sent to and received from HP Labs. We however considered a graph
with the internal and external nodes mixed (as in
\cite{ebel02email}) to demonstrate the effect of a decay on the
spread of email specifically in a power-law graph.

We simulated the spread of an epidemic by selecting a random
initial sender to infect and following the email log containing
120,000 entries involving over 7,000 recipients in the course of a
week. Every time an infective individual was recorded as sending
an email to someone else, they had a constant probability $p$ of
infecting the recipient. Hence individuals who email more often
have a higher probability of infecting. We also assume that an
individual remains infective (willing to transmit a particular
piece of information) for a period of 24 hours.

Next we introduced a decay in the transmission probability $p$ as
$p*d_{ij}^{-1.75}$, where $d_{ij}$ is the distance in the
organizational hierarchy between two individuals. This exponent
roughly corresponds to the decay in similarity between homepages
shown in Figure \ref{disttolikeav}. The decay represents the fact
that individuals closer together in the organizational hierarchy
share more common interests. Individuals have a distance of one to
their immediate superiors and subordinates and to those they share
a superior with. The distance between someone within HP labs and
someone outside of HP labs was set to the maximum hierarchical
distance of 8.

\begin{figure}[tbp]
\begin{center}
\includegraphics[scale=0.4]{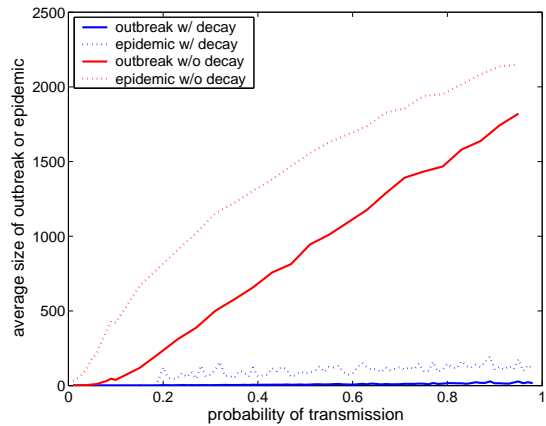}
\end{center}
\caption[Outbreak size as a function of the transmission
probability $p$ ]{Average outbreak and epidemic size as a function
of the transmission probability $p$.} \label{outbreak}
\end{figure}

In figure \ref{outbreak} we show the variation in the average
outbreak size, and the average epidemic size (chosen to be any
outbreak affecting more than 30 individuals). Without decay, the
epidemic threshold falls below $p=0.01$. With decay, the threshold
is set back to $p = 0.20$ and the outbreak epidemic size is
limited to about 50 individuals, even for $p = 1$.

As these results show, the decay of similarity among members of a
social group has strong implications for the propagation of
information among them. In particular, the number of individuals
that a given email message reaches is very small, in contrast to
what one would expect on the basis of a virus epidemic model on a
scale free graph. The implication of this finding is that merely
discovering hubs in a community network is not enough to ensure
that information originating at a particular node will reach a
large fraction of the community. We expect that these findings are
also valid with other means of social communication, such as
verbal exchanges, telephony and instant messenger systems.

\bibliography{hplemail}
\end{document}